%% file: article.tex
\documentclass[10pt]{article}
\input{hicss-packages.tex}

\usepackage{comment}

\newcommand{\substepseparator}{\hspace{1cm}}

\title{Mind the Gap: Reimagining an Interactive Programming Course for the Synchronous Hybrid Classroom}

\author{Christopher M. Poskitt, Kyong Jin Shim, Yi Meng Lau, and Hong Seng Ong \\
  Singapore Management University, Singapore \\
  {\underline{ \{cposkitt, kjshim, ymlau\}@smu.edu.sg}}, \underline{hsong.sg.r@gmail.com} \\}

\date{}

\begin{document}
\maketitle
\begin{abstract} 
    COVID-19 has significantly affected universities, forcing many courses to be delivered entirely online. As countries bring the pandemic under control, a potential way to safely resume some face-to-face teaching is the synchronous hybrid classroom, in which physically and remotely attending students are taught simultaneously. This comes with challenges, however, including the risk that remotely attending students perceive a `gap' between their engagement and that of their physical peers. In this experience report, we describe how an interactive programming course was adapted to hybrid delivery in a way that mitigated this risk. Our solution centred on the use of a professional communication platform---Slack---to equalise participation opportunities and to facilitate peer learning. Furthermore, to mitigate `Zoom fatigue', we implemented a semi-flipped classroom, covering concepts in videos and using shorter lessons to consolidate them. Finally, we critically reflect on the results of a student survey and our own experiences of implementing the solution.
\end{abstract}

\section{Introduction}

The COVID-19 pandemic has had a significant impact on teaching and learning activities at universities around the world, with public safety rules forcing many courses to be delivered entirely online~\cite{DePietro20a,Watermeyer-et_al20a}. While effective remote learning is possible~\cite{Guo-Kim-Rubin14a,Yousef-et_al14a}, students consistently report that they miss face-to-face interactions with their peers and faculty, and find this to be an important part of the learning experience~\cite{Paechter-Maier10a, Ross20a}. In countries that have brought the pandemic under control, a potential way to safely resume some face-to-face teaching is to adopt the \emph{synchronous hybrid classroom}~\cite{Hastie-et_al10a} (also known as the \emph{concurrent classroom}), in which physically and remotely attending students are taught simultaneously (e.g.~using video conferencing in the classroom), taking turns to come to campus.

While hybrid delivery restores many of the benefits of face-to-face teaching, it also creates a number of problems and challenges of its own~\cite{Pullen12a,Detienne-Raes-Depaepe18a,Raes-et_al20a}. First, there is the risk that remotely attending students could perceive a `gap' between their engagement and that of their face-to-face peers. Addressing the needs of both at the same time is very difficult, and a physically present faculty may unintentionally default to interacting more with the physically present students. Second, the format makes peer learning during lessons harder, as some students may never meet in person, and the physically attending students must ensure safe distancing. Third, remotely attending students may be disadvantaged further by the effects of `Zoom fatigue'~\cite{Fosslien-Duffy20a,Wiederhold20a}.

As the spread of COVID-19 in Singapore began to fall under control, our institution selected our front-end web application programming course to be adapted to this synchronous hybrid mode. Prior to the pandemic, our course was delivered fully face-to-face in tiered `U'-shaped seminar rooms (up to 50 students) that helped facilitate a highly interactive class dynamic. Teaching was typically multi-modal, blending presentations, discussions, quizzes, coding demonstrations, and hands-on exercises over weekly 3hr long classes. The key problem we faced was how to adapt this course to the constraints of hybrid delivery in a way that maintained its previous levels of interaction and peer learning, while also ensuring a consistent learning experience between physically and remotely attending students.

In this \emph{experience report}, we describe how our interactive programming course was adapted to simultaneous hybrid delivery in a way that mitigated the main challenges of the format. First, we incorporated the use of a professional communication tool---Slack~\cite{Slack}---making it the principal means of Q\&A, polling, and progress checking during synchronous lessons in order to equalise opportunities for participation. Second, we shifted peer learning online through the use of $\mathtt{\#troubleshooting}$ Slack channels and a Piazza forum~\cite{Piazza}, with contributions incentivised through class participation grades. Finally, we implemented a semi-flipped classroom to reduce fatigue, allowing for shorter and more focused synchronous classes, and freeing up more time for physical/remote consultations instead.

We gauged the effectiveness of our solution through a post-course student survey as well as our own critical reflections as instructors. The results suggest that we successfully adapted much of the face-to-face learning experience to hybrid mode, with students particularly appreciating the brevity of the semi-flipped classroom model as well as the opportunities to participate and seek the opinions of their peers over Slack. We also identified some challenges that remain to be addressed (e.g.~being able to identify overwhelmed or completely lost students without visual cues), and observed the need to introduce clear codes of conducts when providing anonymous communication features. While students and instructors alike prefer fully face-to-face teaching, our solution could be an effective compromise during the recovery phase of the pandemic, or for expanding access to classes more generally, e.g.~through more flexible distance learning.

Our paper is organised as follows. In Section~\ref{sec:context}, we describe the web engineering course that was selected for synchronous hybrid delivery. In Section~\ref{sec:interventions}, we present our hybrid delivery solution, highlighting how it tackles the problems of ensuring equitable participation, maintaining peer learning, and avoiding `Zoom fatigue'. In Section~\ref{sec:survey} we discuss the results of a student survey, before critically reflecting on our own experiences in Section~\ref{sec:reflections} and making some recommendations. Finally, in Section~\ref{sec:related_work} we compare against some key related work, before offering some conclusions in Section~\ref{sec:conclusion}.

\section{Context}\label{sec:context}

This section presents an overview of \emph{IS216: Web Application Development 2}, which was selected by our institution for simultaneous hybrid delivery. We describe how the course was originally intended to be delivered prior to the COVID-19 pandemic, and the concrete constraints of hybrid mode that we were now required to work within.

IS216 is a core course taken in the second year of our undergraduate Information Systems programme, and covers the fundamentals of front-end web programming. The curriculum introduces students to the three principal building blocks of webpages---HTML, CSS, and JavaScript---as well as some modern front-end frameworks, including the Bootstrap framework for responsive design, and the Vue.js framework for reactive user interfaces. IS216 follows on directly from an earlier core course that covers the fundamentals of server-side programming and state management, thus the two courses together give students the ingredients they need to be able to build a full-stack web application.

Instead of traditional lectures, the teaching of IS216 is based on smaller classes. Students bid to register for one section (or cohort) with a capacity of up to 50 students, which then meets once a week for a 3hr slot. Each section is supported by a faculty member, an instructor, as well as a teaching assistant~(TA), i.e.~a senior student who passed a previous iteration of the same course. The term is divided into two halves of seven weeks, and spread across them are a number of different assessments, including a mid-term programming test, a group project, a final written exam, and periodic quizzes that contribute to a class participation grade.

\begin{figure*}[!t]
	\centering
	\includegraphics[width=0.9\linewidth]{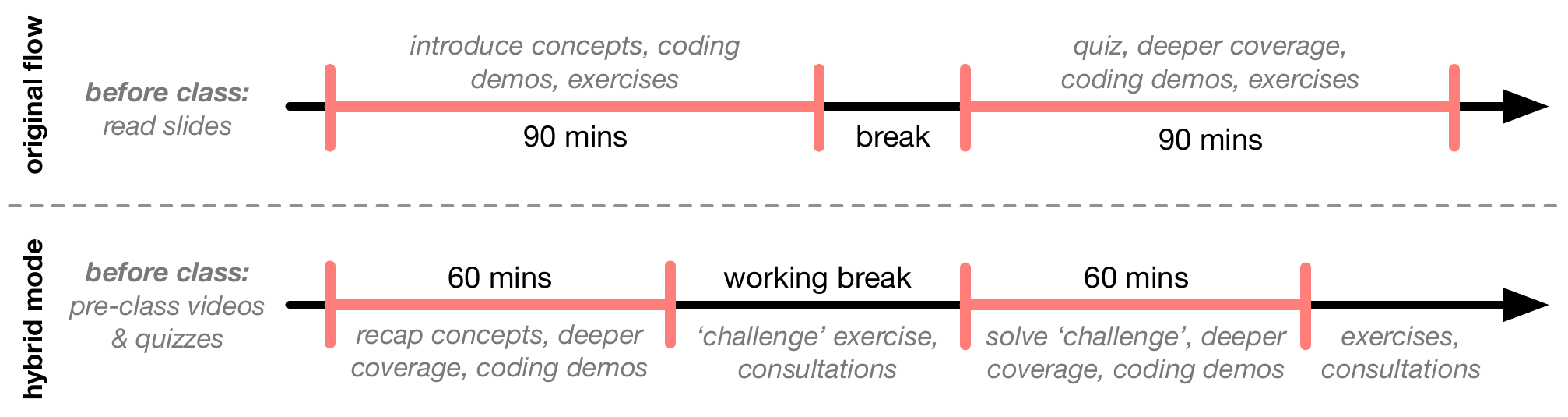}
	\caption{Flow of our weekly synchronous classes before (above) and after (below) switching to hybrid mode}
	\label{fig:class-flow}
\end{figure*}

Prior to the pandemic, IS216 was delivered fully face-to-face, using tiered `U'-shaped seminar rooms that helped facilitate an interactive class dynamic. The lesson flow (Figure~\ref{fig:class-flow}) was typically very multi-modal, in the sense that micro-presentations (e.g.~on new technical concepts) were interleaved with live coding demonstrations, quizzes (e.g.~using student response systems~\cite{Heaslip-Donovan-Cullen14a}), and multiple discussion / Q\&A segments facilitated by the faculty.  Furthermore, students were frequently encouraged to put the new concepts into practice through a number of programming exercises that they were asked to attempt during class. While working on these, students would benefit from peer learning by collaborating with the students sat immediately adjacent, and the teaching staff would rove the classroom to gauge their understanding and offer one-on-one assistance. Such assistance would also be available outside of the class, typically through arranged consultations, email discussions, or communication over messaging platforms such as Telegram.

The 2020-21 iteration of IS216 was scheduled to commence at a time when Singapore's national COVID-19 situation was stabilising. While most other courses were still set to be delivered fully online, given its technical nature, IS216 was selected by our institution to be run as a hybrid class, in which face-to-face and remote students would be taught \emph{simultaneously}, i.e.~by a faculty member physically in the classroom but also running video conferencing software. The rationale was to allow for some face-to-face teaching to resume, but in a way that satisfied national safety guidelines and that would still allow for the full participation of anyone who could not physically attend (e.g.~due to shielding, quarantine orders, or international students affected by border restrictions). Hybrid mode, however, imposed a number of challenging constraints: seminar rooms were restricted to 25 physically attending students; everyone was to be wearing a face mask; nobody was to come within 1m of another person; and students were to be split into A/B teams that alternated between physical/remote attendance every two weeks.

Unfortunately, many of these constraints were not compatible with our traditional mode of delivery. We were concerned that without an effort to adapt, we would see reduced student engagement, lost opportunities for peer support, and simply less interaction in class---especially for those students participating remotely. We were also concerned that hybrid mode could lead to the risk of `slipping' into teaching the physically present students as normal, while neglecting the needs of those who are simultaneously participating from home. That is to say, we saw a risk that remotely attending students could start to feel like second-class citizens of the section, and worried that they would perceive a `gap' between their own experiences and those of their peers in the classroom. Ultimately, these concerns boiled down to the following problem to address:

\begin{center}
\noindent\fbox{%
    \parbox{0.9\linewidth}{%
        \small\emph{How do we adapt interactive programming classes to hybrid delivery, while ensuring a consistent experience between physical and remote students?}
    }
}
\end{center}

\section{Our Interventions}\label{sec:interventions}

In designing a hybrid delivery solution for IS216, we identified the following three sub-problems. First, how do we ensure \emph{equitable participation} and access to support for all of our students, whether attending the classes physically or remotely? Second, how do we continue to maintain a high degree of \emph{peer learning} given the imposed distancing constraints? And finally, how do we adapt the flow of our lessons to ensure that our remote students do not succumb to `Zoom fatigue'?

We present the details of our solution in the following, addressing the three sub-problems in turn. We also present a side-by-side summary in Table~\ref{tbl:differences}.

\begin{table*}[t]
\caption{Comparison of fully face-to-face and hybrid classes} \label{tbl:differences}\vspace{10pt}
\centering\footnotesize\begin{tabular}{l||p{2.2in}|p{2.8in}}
\textbf{Course Aspect} & \textbf{Original Face-to-Face Delivery} & \textbf{Simultaneous Hybrid Delivery} \\
\hline\hline
Assessments & Class participation, group project, \emph{closed} programming test, closed final written exam & Class participation, group project, \emph{open} programming test, closed final written exam \\
\hline
Class duration & 3hrs, once per week & Two 1hr segments (plus consultations), once per week \\
\hline
In-class exercises & Simple programming exercises interleaved in synchronous classes, extra exercises at home & Exercise(s) during `working breaks' and consultation segments, extra exercises at home \\
\hline
In-class discussion & Facilitated by faculty in U-shaped seminar room. Students raise hands to contribute, or just speak & Largely conducted over the section's in-class communication channel on Slack, facilitated by the faculty \\
\hline
In-class support & Faculty, instructor, TA, and peers provide face-to-face support & Faculty focuses on physically present students, instructor focuses on remote ones, and TA assists both where needed \\
\hline
Open consultations & By appointment & During `working breaks', by appointment, or over Slack DMs \\
\hline
Peer-learning & Project groups, or via adjacently-sat students & Project groups, Piazza forum, in-class Slack channels \\
\hline
Pre-class activities & Read class slides & Watch instructor-prepared concept videos, do pre-class quizzes \\
\hline
Progress checking & Faculty, instructor, and TA rove the classroom and informally speak with students & Faculty asks students to `react' on Slack (e.g.~with a `thumbs up' emoji) when they reach certain checkpoints \\
\end{tabular}
\end{table*}

\substepseparator

\noindent\textbf{Ensuring Equitable Participation.} In our original face-to-face mode, faculty were able to facilitate reasonably equal opportunities for students to participate by virtue of the small and intimate `U'-shaped classrooms. Teaching teams were also able to ensure a fairly distributed amount of one-on-one support during exercise segments when they would rove the classroom and speak with students. The shift to hybrid mode, however, with half of the class dialling in over Zoom, meant that there was a very real risk that half the class could become neglected.

To alleviate this problem, our IT Support department installed some Video Conferencing~(VC) equipment in classrooms designated for hybrid teaching. The goal was to allow remote students to feel `closer' to the classroom, by providing a live feed from a ceiling camera and the audio from the room's existing microphones. At the beginning of each class, the faculty was to set up a regular Zoom meeting on their laptops, then use the room's wall panel to have the VC system dial in. After the room `joins' as a participant, the Zoom meeting would be able to share live broadcasts of the faculty's laptop as well as the feed from the ceiling camera. This connection was not just one-way: a remote participant who unmuted themselves and spoke would have their voice broadcast over the physical classroom's speakers, allowing for a live discussion to take place.

While glad of this technological intervention, we were concerned that it was not enough for the hybrid format to be inclusive. First, we suspected that shier remote students would be dissuaded from speaking, given that their voices would be broadcast over the physical room's speakers and recorded. Second, it provided limited means for teaching staff to check up on remote students: Zoom's text chat feature is limited (and difficult to access with our VC set up), and we lacked visual cues as we did not want to mandate webcam usage (to avoid intruding on students' personal spaces or broadcasting socioeconomic differences~\cite{Marquart-Russell20a}). Finally, we were concerned about \emph{our} human nature as faculty: that we would unintentionally default to the `easier' option of focusing our interactions on those physically present, enlarging the participation gap ourselves.

\begin{figure}[!t]
	\centering
	\includegraphics[width=1\linewidth]{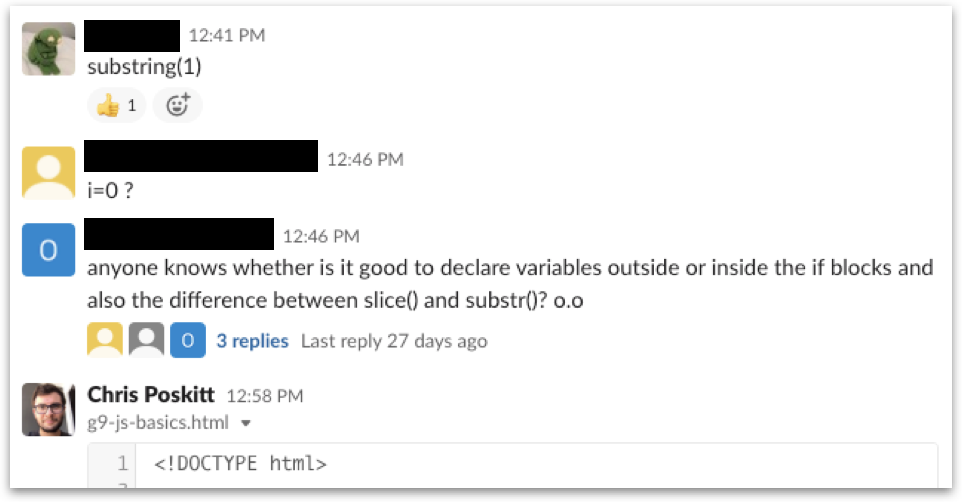}
	\caption{Using Slack during a hybrid class}
	\label{fig:slack}
\end{figure}

To solve this problem, we decided to introduce the use of a professional communication platform, Slack~\cite{Slack}, as a way of equalising participation opportunities between the physical and remote students. In particular, we created a dedicated channel on the platform for each section (e.g.~$\mathtt{\#g8}$, $\mathtt{\#g9}$), then asked \emph{every} student---whether remote or physical---to keep their section's channel open during class. We encouraged every student to use it as a principal means of communication during synchronous class, including asking questions\footnote{Physical students could still ask questions orally, but we gave these equal priority to questions asked in the class Slack channel.}, answering questions (both those of the faculty and their peers), reporting progress, and sharing code snippets. Figure~\ref{fig:slack} contains a screenshot from a synchronous hybrid class: the first two students were answering questions posed orally by the faculty; the third student asked a question of their own, which spurred a thread; then finally, the faculty uploaded the code from a live demonstration.

This in-class Slack channel provided a number of advantages over alternatives such as Zoom's text chat function. First, all students could easily access it (physically attending students are not in the Zoom meeting). Second, we could quickly check-up on students' progress or status by asking them to \emph{react} to questions using emojis, e.g.~``Are you ready to start?'', ``React with a \emph{thumbs up} when you're ready for me go over the exercise'', or even ``React with \emph{A}, \emph{B}, \emph{C}, or \emph{D}'' for a multiple-choice question posed to the class on a slide. Third, when asking a question to the class and encouraging them to type the answer in Slack, the platform indicates when students are typing, allowing faculty to wait in the confidence that an answer is forthcoming. Fourth, the platform comes with a substantial number of useful features and apps, such as the ability to share formatted code snippets, or the Anonymous Bot app~\cite{Anonymous-Bot} to allow students to ask questions they are `shy' about without revealing their identities. Fifth, the discussions are persistent, allowing students to easily review what happened in previous classes. Finally, using Slack provides them with experience in using a communication platform that they are likely to encounter in industry~\cite{Lin-et_al16a}.

We organised communication over Slack and Zoom as follows. During synchronous class, the faculty used two laptops: one for broadcasting/presenting, and the other used solely for handling the Slack channel (so as not to inadvertently broadcast any private Slack messages over Zoom). When posing questions to the class, the faculty would monitor the Slack channel for typing, answers, or reactions as appropriate. While presenting to the class (e.g.~a live coding demonstration), the Slack channel would be left open for students to type questions as they came to mind, which the faculty would attempt to answer at appropriate points of the presentation---if not already answered directly in the channel by the instructor, TA, or other students.

During exercises and breaks, to ensure equitable support, each member of the teaching team would target different parts of the class for one-on-one consultations: the faculty would focus on the physically present students, the instructor on the remotely attending ones, and the TA would support both according to demand. When supporting remote students (whether during or after class), we used Slack Direct Messages (DMs) to discuss and share code, switching to Zoom breakout rooms when questions were too broad to address in text chats.

\substepseparator

\noindent\textbf{Maintaining Peer Learning.} In our original mode of delivery, peer learning was encouraged by the teaching team in two ways. First, through an open group project, in which teams of up to five students were tasked with building a front-end web application that solves a problem of their choice while satisfying a number of technical requirements (e.g.~must call an external API using JavaScript and JSON, must have a mobile-friendly design). Second, through the in-class exercise segments, in which students would work together with their immediately adjacent peers.

In our hybrid solution, the group project was largely unaffected as the bulk of the work was undertaken outside of the classroom. Groups coordinated over messaging platforms, but they could also occasionally meet together off-campus, where they were subject only to national safety restrictions (which allowed gatherings of up to five at the time; same as the group sizes). The only aspect of the project we had to adapt was its assessment, which originally involved a presentation and demo in front of the whole class. We felt that this was no longer practical, as groups were typically split across A/B teams (so could not present together), and we did not want groups that were forced to present remotely to perceive that they were at a disadvantage to groups presenting physically. We also anticipated a mix of technical issues and fatigue if \emph{every} group was asked to present remotely. To solve this, we instead asked each group to submit a short (approx.~10 mins) YouTube video presenting their chosen problem, solution, and demo; we then scheduled short meetings for Q\&A only. We also encouraged teams to deploy their web applications online so that teaching staff and peers alike could easily interact with them asynchronously.

For in-class exercises, however, the A/B split and 1m safe distance rules required us to intervene in order to maintain some peer learning. Our solution was again to leverage communication platforms that could be used by both the physical/remote halves of the class. For example, dedicated channels (e.g.~$\mathtt{\#troubleshooting}$, or section-specific channels as in Figure~\ref{fig:slack}) were set up in Slack, giving students a place to ask and answer each other's questions, with the teaching team jumping in from time-to-time to verify certain answers or fill in any gaps. Where Slack's Anonymous Bot app was enabled, students could also post their questions anonymously, allowing them to seek help from their peers even for questions they worried might be seen as `too simple'.

\begin{figure}[!t]
	\centering
	\includegraphics[width=1\linewidth]{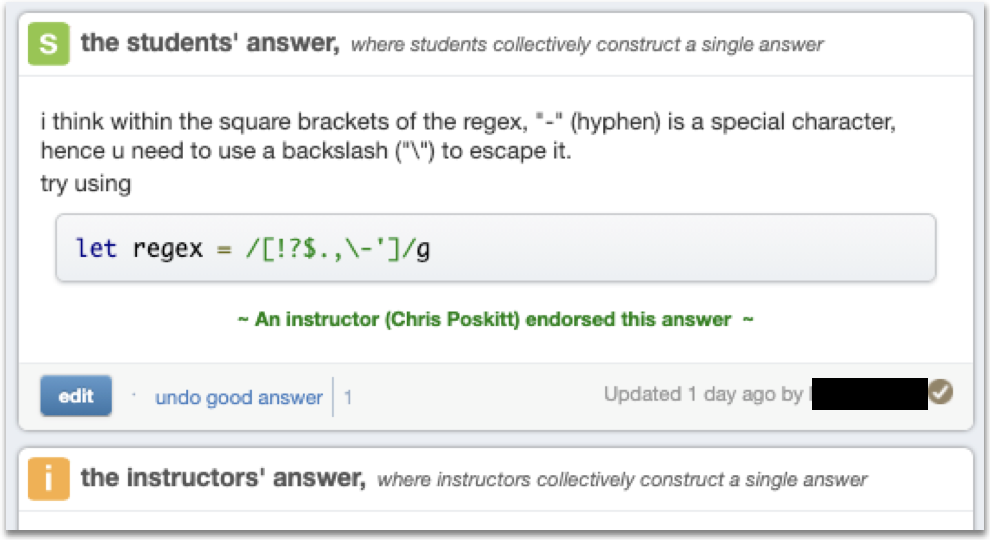}
	\caption{Shifting peer learning online using Piazza}
	\label{fig:piazza}
\end{figure}

In addition to facilitating this peer support in Slack, we also trialled the Piazza Q\&A platform~\cite{Piazza} in some sections to allow students to ask and answer longer technical questions outside of class and with different degrees of anonymity (e.g. anonymous to classmates, but not faculty). This platform was explicitly designed to support peer learning: students can collaboratively answer questions, and instructors can then endorse those answers to instil extra confidence in their accuracy (see, for example, an endorsed answer in Figure~\ref{fig:piazza}). To incentivise students into supporting their peers on Slack and Piazza, we holistically considered strong contributions to peer learning across these platforms towards their `class participation' grade components.

\substepseparator

\noindent\textbf{Avoiding `Zoom Fatigue'.} In our original face-to-face delivery, most of the learning was designed to take place in our weekly 3hr classes (typically split into two 1hr30min halves), which interleaved micro-presentations for introducing concepts, quizzes for improving comprehension, and hands-on exercises for applying them to new problems (Figure~\ref{fig:class-flow}). We expected students to consolidate their learning outside of class (e.g.~revision, extra exercises), but generally did not expect much pre-class preparation other than skimming through slides in advance. While our weekly classes were no doubt long, their interactive and multi-modal nature---not to mention everyone's physical presence---helped ensure a high level of engagement. We were very concerned that this engagement would not translate to hybrid mode, especially for the students participating over VC software, which can be fatigue-inducing~\cite{Fosslien-Duffy20a,Wiederhold20a}.

\begin{figure}[!t]
	\centering
	\includegraphics[width=1\linewidth]{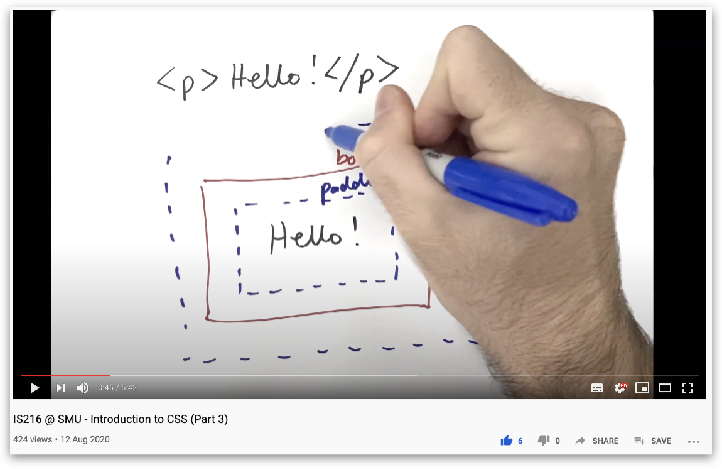}
	\includegraphics[width=1\linewidth]{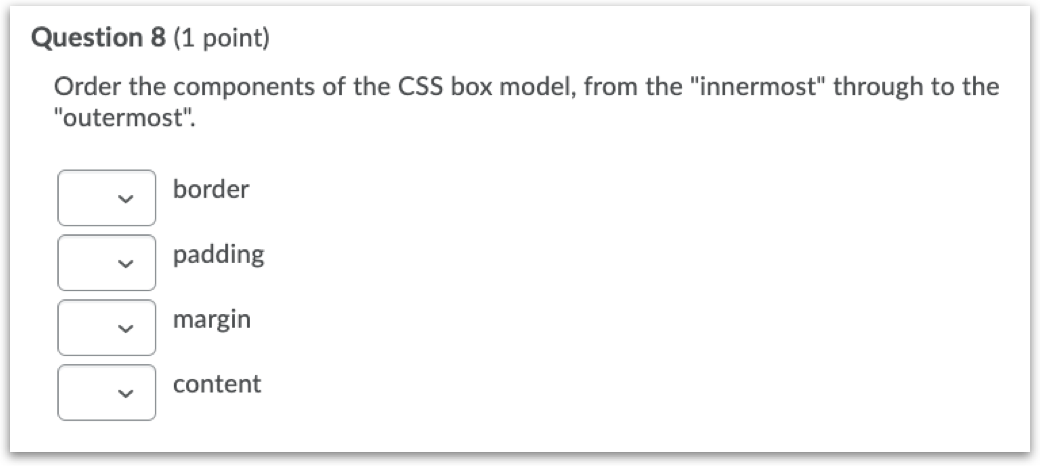}
	\caption{Pre-class video and aligned quiz question}
	\label{fig:preclass-video-quiz}
\end{figure}

To solve this problem, we designed our hybrid solution to incorporate elements of the flipped classroom~\cite{Lage-Platt-Treglia00a,Mok14a,Maher-et_al15a}. In particular, instead of covering concepts in-class, we shifted them out into \emph{pre-class activities}, allowing us to run shorter synchronous classes that focused on applying the concepts and discussing deeper technical issues. Our pre-class activities each week consisted of 2-3 tailor-made YouTube videos that covered the main technical concepts using high-level examples and narrated coding screencasts~\cite{PreClassVideos,PreClassVideos2}. For example, prior to the CSS lesson, our videos covered CSS rules, classes, and the box model (Figure~\ref{fig:preclass-video-quiz}), demonstrating each topic with a simple and independent worked example. We designed each video to be short and digestible, in line with the recommendations of recent studies (e.g.~\cite{Guo-Kim-Rubin14a,Long-Logan-Waugh16a}). Furthermore, to ensure that this learning was not passive, we prepared pre-class quizzes with questions that were constructively aligned~\cite{Biggs96a} to the learning outcomes of the videos (see again Figure~\ref{fig:preclass-video-quiz}). The intention was to give an immediate sense of progress and ensure all students would be primed for the synchronous classes. To motivate students to prepare for and complete these quizzes, we considered them towards their class participation grade components.

With concepts covered beforehand, we were able to reduce the length of our synchronous classes to approximately two 60 minute segments, with a longer `working break' between them, as well as free time for exercises/consultations at the end (Figure~\ref{fig:class-flow}). Our two synchronous segments followed the topics of the videos, briefly recapping each concept before diving into deeper technical issues, typically using interactive live coding demonstrations. While multi-modal like our original face-to-face delivery, we adapted our handling of in-class exercises: during synchronous broadcasts, we focused on simpler polling exercises (e.g.~voting for an option in Slack), shifting the hands-on programming exercises to working breaks and in-class consultation time. This streamlined the logistics of providing one-on-one support, and minimised the risk of exercises artificially extending the lengths of the two Zoom broadcasts (potentially risking fatigue). The hands-on exercises set during working breaks would typically be discussed and solved together at the start of the second synchronous segment.

\section{Student Survey Results}\label{sec:survey}

To obtain some reflections from our students, we designed a short and optional survey which we sent to them after the completion of all assessments and grading. Our survey consisted of several statements (e.g.~``I actively ask questions during class'', ``I am able to consult the teaching team during class''), and we asked students to indicate their level of agreement using a five-point Likert scale, first with respect to their previous (prerequisite) programming course that was delivered fully face-to-face, and second with respect to our synchronous hybrid delivery of IS216. Following this, we asked questions regarding the utility of our technological interventions in hybrid mode. Finally, we asked open-ended questions about what worked well and what could be improved about our hybrid solution.

\begin{table}[!t]
    \caption{Survey: average agreement (5-point Likert scale) with respect to original and hybrid classes}
    \label{tab:survey_results1}
    \vspace{10pt}\centering\footnotesize
    \begin{tabular}{p{1.7in}||c|c}
\textbf{Statements} &	\textbf{Fully F2F} &	\textbf{Hybrid} \\
\hline\hline
I actively ask questions publicly during class &	2.97 &	2.98 \\
I actively ask questions privately during class	& 3.80 &	3.72 \\
I actively answer questions posed by the teaching team during class &	3.38 &	3.29 \\
I actively answer questions posed by peers during class	& 3.08	& 3.10 \\
I am able to learn from peers during class	& 3.91	& 3.76 \\
I can stay actively engaged for the full length of the synchronous class	& 3.69	& 3.62 \\
I am able to get sufficient break(s) during class	& 4.14	& 4.22 \\
I am able to consult the teaching team during class	& 4.18	& 4.11 \\
I am able to get sufficient hands-on coding practice during class	& 4.12	& 4.08 \\
I am able to get sufficient hands-on coding practice outside of class	& 4.21	& 4.22 \\
I am able to consult the teaching team outside of class	& 4.33	& 4.31 \\
I am able to learn from peers outside of class	& 3.97	& 3.97 \\
Weekly/bi-weekly quizzes help me review the concepts	& 4.30	& 4.31 
    \end{tabular}
\end{table}

\substepseparator

\begin{figure}[!t]
    \centering
    \includegraphics[width=\linewidth]{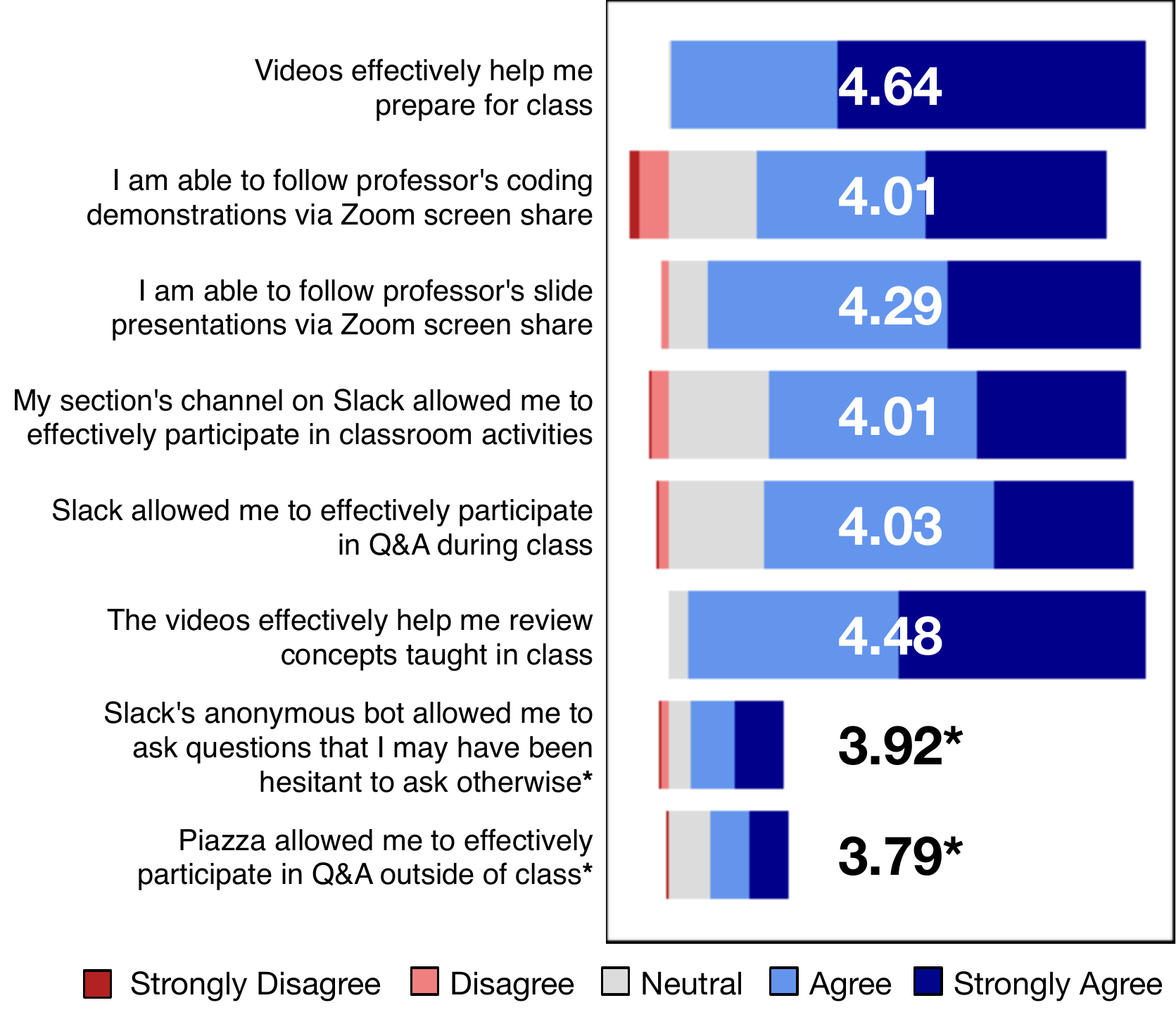}
    \caption{Survey: average agreement regarding our technological interventions}
    \label{fig:survey_results2}
\end{figure}

\noindent\textbf{Results.} Our survey elicited 147 responses from the five sections surveyed (managed by two faculty), indicating a response rate of approximately $60\%$. The results are given in Table~\ref{tab:survey_results1} and Figure~\ref{fig:survey_results2}, and are based on a 5-point Likert scale from `Strongly Disagree'~(1) to `Strongly Agree'~(5). Questions marked ($\ast$) concern technology that was only trialled in two of the sections (38 students).

Across our statements regarding engagement and access to support (Table~\ref{tab:survey_results1}), we found the responses fairly consistent between fully face-to-face and hybrid mode. We are satisfied with the overall consistently of these scores: they reflect the enduring popularity of regular face-to-face classes, but suggest that our hybrid solution reasonably mitigated the challenges we anticipated and successfully translated much of our original classroom experience. We note a few small reductions, e.g.~from 3.91 to 3.76 for ``I am able to learn from peers during class'', suggesting that we could optimise further (e.g.~by assigning peer buddies).

We also saw high levels of satisfaction with our technological interventions (Figure~\ref{fig:survey_results2}): $99\%$ of students agreed or strongly agreed that our videos helped prepare them for class (and for $96\%$, helped them review concepts); $75\%$ agreed or strongly agreed that Slack facilitated class participation; and $77\%$ agreed or strongly agreed that Slack facilitated effective Q\&A during class (with some sections reporting up to $97\%$ agreement). The questions regarding `Anonymous Bot' and Piazza also elicited positive responses, but have a lower average score due to relatively more `Neutral' responses from students who did not use them (only three and one responses respectively were negative).

In written responses, many students highlighted that they ``don't focus as well'' in general during online lessons, so appreciated the ``brevity of [our] online classes''; that it was ``easier to focus for the entirety of the lesson'', and gave ``more time to practice coding on their own''. Several students commented positively on the pre-class materials and said that they allowed them to prepare effectively for class. Several students also commented positively on the use of Slack, in particular, that ``everyone was given equal chance to engage either in-class or at home'', and that ``[they] can get opinions from [their] classmates'' using it. Students also ``[liked] how there are lots of platforms to ask questions'', including Slack, Piazza, and regular email. While some students appreciated the possibility of engaging in Q\&A anonymously, one student reflected that it might be better to remove this so that they grow to ``bravely ask questions instead of doubting [their] own questions''. Finally, some students highlighted positive aspects of hybrid mode that are not usually possible in fully face-to-face classes: in particular, the ability to use a second monitor at home to code more efficiently (e.g.~by using it to display the faculty's live lecture, while coding simultaneously on a laptop).

\substepseparator

\noindent\textbf{Limitations \& Open Questions.} We are encouraged by the results of the survey, which increase our confidence that much of the original classroom learning experience was successfully translated into synchronous hybrid mode. The students' written responses particularly encouraged us that our design decisions achieved their intended outcomes. It is important to remark, however, that the urgency of designing/implementing our solution precluded a more formal experimental design, and thus there are potential threats to validity that should be investigated in future work. For example, in Table~\ref{tab:survey_results1}, students are comparing our hybrid mode against the fully face-to-face delivery of a different (but closely related) course---the pandemic made it impossible to make this comparison within IS216 only.

There are a number of interesting open research questions regarding our hybrid solution that would merit some experimental evaluation in the future. For example, can our solution be generalised to other programming courses, or even other kinds of information systems and computing courses? How effective is our solution at equalising opportunities for peer learning, and what instruments should be used to measure this? And while our students' perceptions of hybrid mode are positive, how effective is it (in comparison to fully face-to-face) at helping them achieve various learning objectives? Are there differences in effectiveness between objectives from different levels of Bloom's taxonomy (e.g.~knowledge and comprehension versus synthesis and evaluation)?

\section{Critical Reflections}\label{sec:reflections}

In this section, we critically reflect on our experiences of implementing this hybrid solution, and whether it was able to achieve its aims. The general consensus of the teaching team is that our hybrid solution successfully translated many of the original face-to-face features (e.g.~Q\&A, interactivity), but fell short on a few others (e.g.~identifying struggling students) that were inherently difficult in the format. We also identified a few features of our hybrid solution that we will \emph{retain} once we are able to return to fully face-to-face classes, e.g.~the use of Slack as a communication platform.

We found that Slack was an effective platform for encouraging interaction and communication during hybrid lessons, both for students attending physically and those at home. To our initial surprise, we sometimes found students more interactive on Slack than they had been in previous fully face-to-face courses, and that many of our shier students were more willing to answer questions using it. We were encouraged by this observation, and would be keen to continue using an in-class text channel---even in fully face-to-face classes---to maintain an inclusive learning environment. At the same time, we feel that this should not be to the total exclusion of asking/answering questions orally, given that university is the ideal setting to work on soft skills such as confidence in speaking publicly. While Slack's Anonymous Bot was occasionally used by students to ask meaningful questions, we observed several usages of it in jest, suggesting that proper codes of conducts should be provided in future classes. Having said that, as instructors, we did not conclude that the bot was a `must have' feature of our solution, especially as students seemed comfortable to ask questions privately over Slack DMs or Piazza.

Slack was useful for quickly gauging the progress and understanding of students (especially remote ones), and we frequently used it for quick multiple choice questions and checking progress on exercises during the working breaks. It was less useful, however, for identifying individual students who were overwhelmed or completely lost: the typical symptom of this---silence on Slack---was also a symptom of those who were finding the course straightforward (e.g.~students with prior web development experience). In fully face-to-face classes these students can be distinguished by visual cues, or quick conversations, which in hybrid mode could only happen on certain weeks. Slack was also hard work for us to manage: we required a two-laptop solution to maintain privacy, imposing a lot of context switching. At first, we would try to address incoming questions as they were posted on Slack, but this disturbed our trail of thought. Later, we shifted to checking for questions at more natural points of the presentations, with the instructor and TA helping to answer questions directly in the channel when the faculty was busy. Finally, we found Slack DMs to be the students' preferred method of remote one-on-one communication (very few sent emails or requested video consultations): students seemed to appreciate the ability to upload code snippets and digest our detailed written replies at their own pace.

We found the use of Slack and Piazza to be effective substitutes for the peer learning that previously took place during class. In Slack channels, we found that there was a reasonable number of students who were very engaged and willing to answer the questions of their peers. For sections using Piazza, we observed an even broader participation (perhaps due to the platform's anonymity). With more than 150 threads posted over the term, the platform saw a steady flow of usage ($\sim$3-10 threads per week), which increased in more technical weeks (JavaScript and Vue), and then rocketed up in the week before the final exam. Our sense was that students benefited by having the option of different communication platforms to suit their learning styles. As instructors, we also benefited from public Q\&A discussions (whether on Slack or Piazza), as it reduced the number of repeated questions that we received. 

From our perspective, the switch to a semi-flipped classroom achieved its goals: our pre-class activities saw high levels of engagement (more than 8000 video views; most students completed every quiz), allowing for our shorter synchronous class time to be used more meaningfully. Based on informal Slack check-ups, students largely stayed engaged for the full length of our core 60 minute segments. However, we felt that our adaptation of in-class exercises could be improved: while students engaged with exercises during the working breaks, we observed that some of our remote students disconnected as soon as the exercises/consultations part at the end began. We believe this can be addressed by breaking the class up further into even shorter synchronous segments (e.g.~3--4), thus increasing the number of working breaks and opportunities to consolidate the learning outcomes of exercises. As a final reflection, we must remark that the pre-class materials imposed an enormous amount of additional work on us. While we can re-use them in future runs of IS216 (even fully face-to-face ones), we underestimated the amount of time it would take to prepare high-quality videos and quizzes.

\section{Related Work}\label{sec:related_work}

Menzies and Zarb~\cite{Menzies-Zarb20a} recently reported their experiences of using Slack for \emph{asynchronous} support in four modules of study. In a pre-course survey, they found that students has previously used several online messaging platforms to communicate about their studies, but only 31 out of 144 had previously used Slack. In a post-course survey, after students gained some experience in Slack, they highlighted several benefits, especially the fact that it had all contacts (students and staff) together on one platform---similar to IS216. The authors made a number of recommendations for practitioners, including that a clear code of conduct for communicating on Slack should be provided. We agree with this recommendation, especially after observing some misuse of the Anonymous Bot plugin that we trialled in some sections.

Krusche and Seitz~\cite{Krusche-Seitz19a} also used Slack to increase interactivity, but did so within the context of an asynchronous large-scale software engineering MOOC. Students were encouraged to use a dedicated Slack workspace for contacting TAs, and channels were provided for posting questions publicly (promoting peer learning). In a post-course survey, 57\% of students agreed that Slack was preferable over the traditional discussion forums of MOOCs.

Barr et al.~\cite{Barr-et_al20a} describe their experiences of shifting intensive software engineering modules online due to COVID-19, including a web engineering module. Though \emph{fully} online (rather than hybrid), some of their interventions share similarities with ours. For example, their online synchronous segments were broken up into smaller chunks after students experienced fatigue. Furthermore, incentives were provided to encourage students to provide peer support within smaller assigned groups. Assigning peer groups could complement our solution which tended to focus on `whole class' peer support (i.e.~in Slack and Piazza).

Similar to us, Triyason et al.~\cite{Triyason-Tassanaviboon-Kanthamanon20a}, implemented a hybrid classroom in response to COVID-19. Their paper focuses on how their institution solved the problem of choosing a camera and audio system, as well as a platform for video conferencing. After interviewing three instructors, 15 design guidelines were proposed which were analysed against the features of Meet, WebEx, Zoom, and Teams. Some of our requirements (e.g.~the need for breakout rooms) overlapped with their guidelines, although they did not focus on delivery challenges such as ensuring equitable participation.

Pullen~\cite{Pullen12a} critically reflects on the use of synchronous hybrid learning in a Master of Science in Computer Science programme. Several positive aspects are highlighted, including that the model allows for more flexible distance education, and the possibility of combining multiple sections into one. Negative aspects include the classroom equipment requirements, the possibility of disruptions (e.g.~due to network failure), and faculty `technophobia'. Pullen reflects that once faculty are familiar with the technology, they can quickly `go online' without much additional overhead by using their existing slides. However, he also highlights that additional pedagogical interventions (such as those we have explored) may be needed to promote student interaction.

While our institution was able to conduct in-person final exams as normal, this may not be possible in other situations. One solution for programming courses would be to take remote exams on the students' own devices, possibly using a lockdown browser~\cite{Kurniawan-Lee-Poskitt20a}.

\section{Conclusion}\label{sec:conclusion}

In this experience report, we presented a solution for adapting an interactive web programming course to a synchronous hybrid classroom. First, we incorporated the use of a professional communication platform (Slack) during synchronous lessons to equalise the opportunities for participation. Second, we shifted peer learning online, through the use of in-class Slack channels and a Piazza Q\&A forum, encouraging contributions through class participation grades. Finally, we adjusted the class flow to mitigate `Zoom fatigue', shifting concepts to pre-class videos/quizzes, shortening the synchronous segments, and using working breaks and consultations to facilitate more one-on-one support.

The results of a student survey suggest we successfully adapted much of the face-to-face learning experience to hybrid mode. Students and instructors alike observed that our technological interventions (e.g.~Slack, Piazza) helped equalise opportunities for participation/support, and that a semi-flipped classroom effectively mitigated the risks of fatigue for remote students. While fully face-to-face teaching maintains an enduring popularity, our hybrid solution might be an effective compromise during the recovery phase of the pandemic. Furthermore, it might be effective for expanding access to university courses more generally, e.g.~through more flexible distance learning.

\section*{Acknowledgements}
We are grateful to several colleagues for helpful discussions on the ideas presented, especially Swapna Gottipati, and the participants at the Centre for Teaching Excellence's `Mixed Method Teaching' seminar.


\bibliographystyle{ieeetr}
\bibliography{references}

\end{document}

%% file: hicss-packages.tex
\usepackage[letterpaper]{geometry}
\usepackage{hicss}
\usepackage{times}
\usepackage[none]{hyphenat}
\usepackage{url}
\usepackage{latexsym}
\usepackage{indentfirst}
\usepackage{graphicx}
\graphicspath{{images/}}